# Study of phonon transport across Si/Ge interfaces using Full-Band phonon Monte Carlo simulation

N. D. Le[1], B. Davier[1,2], N. Izitounene[1], P. Dollfus[1], J. Saint-Martin[1]

[1] Université Paris-Saclay, CNRS, Centre de Nanosciences et de Nanotechnologies, 91120, Palaiseau, France

[2] Department of Mechanical Engineering, The University of Tokyo, Tokyo 113-856, Japan

*Abstract*
A Full Band Monte Carlo simulator has been developed to consider phonon transmission across interfaces disposed perpendicularly to the heat flux. This solver of the Boltzmann transport equation does not require any assumption on the shape the phonon distribution and can naturally consider all phonon transport regimes from the diffusive to the fully ballistic regime. This simulator is used to study single and double Si/Ge heterostructures from the micrometer scale down to the nanometer scale, i.e. in all phonon transport regime from fully diffusive to ballistic. A methodology to determine the thermal conductivity at thermal interfaces is presented. It is also shown that the different transport regimes are correlated to different spectral contributions of the phonon modes to the heat flux along the devices. This local indicator of the transport regime gives new insights into the out-of-equilibrium phonon transport near the interfaces.

## I    Introduction

The management of heat transfer at the nanoscale is crucial for the design and optimization of devices, especially in thermoelectrics [1] and nanoelectronics [2]. However, common macroscopic models such as heat Fourier's formalism [3] that assumes local equilibrium are no longer relevant in nanodevices shorter than the mean free path of phonons [4]. In complex nanostructures, the phonon distributions may thus significantly differ from the Bose Einstein equilibrium statistics, especially if thermal interfaces are likely to play a role in the heat transfer.

Several analytical models are able to capture the out of equilibrium transport regime of heat in homogenous systems [5]–[8]. Beyond this standard case, the modeling of interface thermal resistance, also called the Kapitza's resistance [9], is mainly based on either the Acoustic Mismatch Model (AMM) [10], the Diffusive Mismatch Model (DMM) [11] or their extensions [12], [13][14], [15] [16]. It is only recently that an analytical modeling of heat transfer in heterostructure based on a set of a few parameters relevant at the nanometer scale has been presented [17].

Nevertheless, the most versatile approaches to investigate the thermal transfers at the nanoscale are the use of advanced numerical simulations. Among them, Molecular Dynamics (MD) simulations that calculate the classical trajectory of atoms have been widely applied [18]–[20]. These atomistic and time dependent simulations naturally consider the anharmonicity of phonons via semi-empirical parameters that can now be derived from *ab-initio* methods [21], [22]. However, the application of MD remains limited to systems containing thousands of atoms, i.e. with a typical size in the order of tens of nanometers. The Non Equilibrium Green's Functions framework [23]–[25] is another approach that is able to consider mechanical waves at the nanoscale and thus their possible reflection [24], [26]. In its common form, this method is not time dependent but makes it possible to investigate systems larger than those tractable with MD. However, the implementation of the anharmonicity of the phonon dispersion in this formalism severely increases the computational burden [27].

The stochastic particle Monte Carlo (MC) method [28]–[30] is an efficient and accurate method to solve the time dependent Boltzmann's Transport Equation (BTE) much beyond the linear approximation. This semi-classical framework is relevant from the nano to the microscale even in complex geometries in which thermal interfaces are involved [31]–[33]. The phonon dispersion can be either simplified using a parabolic approximation or accurately described using a full band description [34] capturing the modal contributions [35]. Moreover, accurate descriptions of scattering mechanisms and in particular phonon/phonon scattering can be implemented at the particle level considering scattering rates that can be either semi-empirical or derived from ab-initio methods [36] [37].

In the present work, phonon transport in heterostructures have been investigated in all phonon transport regimes. To capture the phonon mode mismatch at the interface and the contribution of each mode especially at high energy, an accurate description of the material, i.e. of the phonon dispersion and the scattering mechanisms throughout the 3D Brillouin zone, is of great importance. Accordingly, Full-Band MC simulation has been performed.

The implementation of the interface modelling is detailed in section II. Results obtained from MC simulation in both simple and double Si/Ge heterostructures are presented and discussed in section III. The spectral contributions of the phonon modes to the heat flux along the devices is particularly studied.

## II  Modeling of heterostructures

### II.1.  Definitions based on hemispherical temperature concept

In accordance with our formalism developed in Ref. [17], our definitions of the effective/ballistic thermal conductivities $\kappa_{effective}$ / $\kappa_{ballistic}$, the total/ballistic thermal conductances and the interface thermal conductance $G_{int}$ which are all related to these hemispherical temperatures are reminded in this section.

Hemispherical (or directional) temperatures $T^+/T^-$ are defined, even in out-of-equilibrium conditions, as the temperature of phonon sub-populations with a positive/negative (oriented toward the cold/hot thermostat, respectively) giving the same energy density $E$ as that obtained from an equilibrium distribution. Thus, it satisfies the following relationship:

$$E(T^+/T^-) = \sum_{state\ s,\ v_x^s>0/v_x^s<0} \hbar\omega_s f_{BE}(\omega_s, T^+/T^-) \qquad (1)$$

where $f_{BE}$ is the Bose-Einstein distribution function and $\omega_s$ the phonon angular frequency of state $s$. It should be mentioned that the most common approach consists in using a local pseudo temperature considering the entire phonon population [34].

In a 1D homogeneous system of length $L$ in contact with a hot thermostat at a hot temperature $T_H$ and a cold one at a cold temperature $T_C$, the thermal flux Q can be expressed (in all transport regimes) by using these different formula:

$$Q = G_{total} \cdot \Delta T_{contact} = \kappa_{effective} \frac{\Delta T_{contact}}{L} = \kappa_{ballistic} \frac{\Delta T_{local}}{L} = G_{ballistic} \Delta T_{local} \quad (2)$$

where the two temperature differences are defined as follows:

$$\Delta T_{local}(x) = T^+(x) - T^-(x)$$

$$\Delta T_{contacts} = T_H - T_C.$$

For a homogeneous structure in which a temperature bias $\Delta T_{contacts}$ is applied and working at (local pseudo) temperature $\bar{T}$, the thermal conductivity $\kappa_{effective}$ is given by:

$$\kappa_{effective} = \frac{Q}{\Delta T_{contacts}} \cdot L = \frac{\Omega}{(2\pi)^3} \sum_{state\ s} \hbar \omega_s |v_{s,x}| \lambda_{mfp,s} \cdot \frac{\partial f_{BE}}{\partial T}(\omega_s, \bar{T}) \quad (3)$$

where $\Omega$ is the volume of the first Brillouin zone, $\lambda_{mfp,s}$ is the mean free path of phonon mode $s$ and $v_{s,x}$ is the phonon group velocity of the state $s$ along the $x$-direction that is the heat transport direction. The associated conductance $G$ is given by $G = \frac{\kappa_{effective}}{L}$. In long devices, the effective conductivity $\kappa_{effective}$ tends to the standard (or diffusive) thermal conductivity $\kappa_{diffusive}$.

In a short or ballistic device $\lambda_{mfp,s} = L/2$ and thus the ballistic conductivity $\kappa_{bal}$ is given by:

$$\kappa_{ballistic} = \frac{L}{2} \frac{\Omega}{(2\pi)^3} \sum_{state\ s} \hbar \omega_s |v_{s,x}| \frac{\partial f_{BE}}{\partial T}(\omega_s, \bar{T}) \quad (4)$$

In heterostructures, interfaces have their own thermal conductance. The interface thermal conductance is here defined as follows [16]:

$$G_{int} = \frac{Q}{\Delta T^I} = \frac{Q}{T^+(x-\varepsilon) - T^-(x+\varepsilon)} \quad (5)$$

where the local temperature at interface $\Delta T^I = T^+(x-\varepsilon) - T^-(x+\varepsilon)$ is the difference between the hemispherical temperatures on each side of the interface.

It should be noted that all thermal parameters (conductivities and interface conductances) are defined using hemispherical temperatures. Indeed, the temperature in a contact is a hemispherical temperature as $T_H = T^+$ and $T_C = T^-$. More details about this formalism can be found in [17].

Finally, Eq. (3) and (4) are used to compute the semi-analytical results presented in this paper. The definitions of Eq. (2) are used to compute the conductivities $\kappa_{effective}$ and $\kappa_{effective}$ from the MC results while Eq. (5) is used for computing $G_{int}$.

## II.2. Full-Band description of phonons

In this study, two materials have been investigated: cubic Si and cubic Ge. Performing Full-Band Monte Carlo simulations requires the prior knowledge of the phonon dispersion, phonon group velocity and phonon scattering rates in the full Brillouin zone. The phonon dispersion in both cubic Si and Ge have been calculated by using the Adiabatic Bond Charge Model (ABCM)[38] [39]. The first Brillouin zone was discretized into a set of 29791 (31×31×31) wave vectors. The phonon band structure in cubic Ge is shown in Fig. 1a. The details about the fitting method and phonon dispersion in cubic Si are given in Ref. [40].

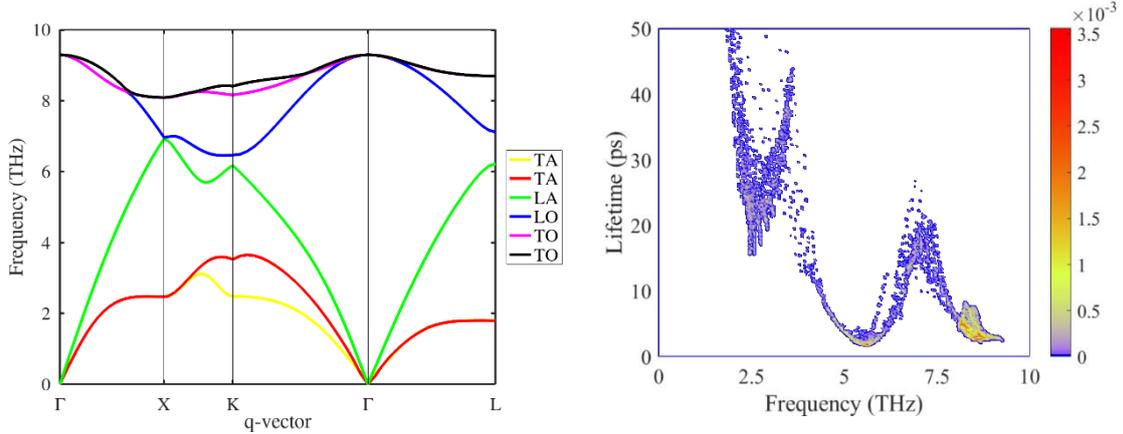

*Fig. 1(a).Phonon dispersion in cubic Ge fitted from ABCM method. (b) Cartography of phonon lifetime in cubic Ge. The colorbar indicates the density of states.*

The phonon-phonon scattering rates in cubic Si have been computed by using the Density Functional Theory (DFT) as previously done in [37], [41]. For Ge, the same spectral distribution as in Si is used (as Si and Ge have the same crystallographic structure) but a proportionality factor on their amplitude was applied to fit the experimental thermal conductivity of Ge at ambient temperature. The spectral phonon lifetimes in Ge used in this work, i.e. the inverse of the phonon scattering rate, are shown in Fig. 1.b.

### II.3. Monte Carlo simulation for phonon

The particle Monte Carlo (MC) method is a stochastic method to solve the time dependent Boltzmann Transport Equation (BTE) without any assumption on the shape of the phonon distribution functions. The Full-Band phonon dispersion and scattering rates of the materials were used as inputs for the Monte Carlo simulation.

Details of the Monte Carlo simulation employed in this work can be found in Ref [34]. It consists of a stochastic particle Monte Carlo algorithm for phonons in which the implementation of phonon-phonon scattering mechanisms is based on the two-phonon approach developed in [29]. Energy conservation is ensured because all simulated particles represent phonon packets that carry the same energy but may differ in the number of phonons in the packets. Of course, all particles can have different positions, wave vectors and modes. In addition, a variance reduction method is used to speed up the simulations [41].

At each simulation time step, the values of the local pseudo-temperature at each position in the structure are updated according to the local phonon distributions. This update is mandatory as this pseudo-temperature is used to find the final phonon state after a phonon-phonon scattering event.

In addition to the contacts with thermostats (which are assumed to be perfect black bodies i.e. perfect emitters and absorbers [34]) two kinds of boundaries are considered in the MC simulation: semi-infinite and rough boundaries. Semi-infinite boundaries are defined by using periodic boundary conditions. In that purpose a couple of opposite external faces (i.e. that are located face to face) must be defined a priori. At these boundaries, when a particle exits from one face a particle with exactly the same properties is created on the opposite (and associated) face.

In the case of a rough boundary, when a phonon collides with the boundary, two kinds of elastic reflection can occur: specular or diffusive reflection. The probability of a specular reflection for an incident phonon of wave vector $\boldsymbol{q_0}$ and incident angle $\theta_0$ is provided by the Soffer's model [42]:

$$p_{specular}(|\boldsymbol{q_0}|, \cos\theta_0) = \exp(-(2\cos\theta_0 \Delta |\boldsymbol{q_0}|)^2) \qquad (6)$$

where $\Delta$ is an empirical parameter characterizing the roughness of the interface. Here, the value of $\Delta$ in both Ge and Si is fixed to 0.5 nm. It should be noted that such a high roughness parameter leads to a quasi-diffusive surface for which the specular reflections are unlikely [34].

In the case of specular reflection, the sign of the component of the reflected phonon wave vector normal to the boundary is reversed, i.e. $q_{j',\perp} = -q_{j,\perp}$, which is not trivial in the case of a Full-Band description if the reflection surface is not oriented along a high symmetry plane of the material. In contrast, the component parallel to the boundary is conserved, i.e. $q_{j',//} = q_{j,//}$. It should be mentioned that as a specular reflection has no effect on the heat flux parallel to the interface, it can also be used to model semi-infinite boundaries. In the case of a diffusive reflection (with a probability of $1 - p_{specular}$.), the reflected phonon wave vector is randomly selected on an iso-energy surface that guarantees both the phonon energy conservation and the orientation condition, i.e. $v_{j',\perp} v_{j,\perp} < 0$ for a reflection. To avoid non-physical accumulation of phonons near the rough boundary, the probability of the final state among the iso-energy states is non-uniform. Indeed, according to the Lambert's law, the probability of the state j' with a reflected angle $\theta_r$ is proportional to the component of the group velocity normal to the boundary [34]:

$$p_{diffusive}(\mathbf{q}_{j'}, \theta_r) \propto |\mathbf{v}_{j'}(\mathbf{q}_{j'})| \cos\theta_r \qquad (7)$$

II.4. Heterojunction modeling within Monte Carlo simulation

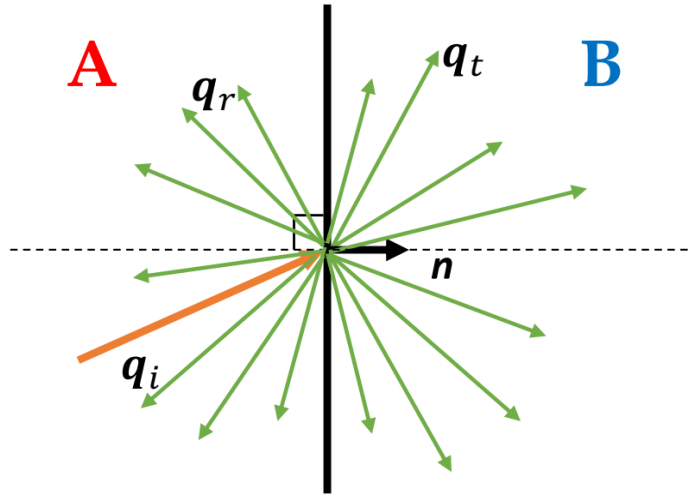

*Fig. 2. Schema of phonon diffusion mechanisms at an interface*

As schematized in Fig. 2, the interface between materials A and B is modeled by using the Full band version of the Diffusive Mismatch Model (DMM) [12]. Even if this approach does not capture all the complex physics of heat transfer at an interface [43], in particular inelastic phenomena are neglected, this approach is easy to implement and provides reasonable prediction of interface thermal conductance at ambient temperature [44] [45]. In DMM it is assumed that all transmitted or reflected phonons have undergone a diffusive scattering at the interface (no specular reflection occurs) and the interface scattering is assumed to be elastic. For an incident phonon of wave vector $\mathbf{q}_j$ and angular frequency $\omega$ colliding with this interface from the A-side, the spectral transmission probability $T_{A \to B}(\omega)$ is given by [44]:

$$T_{A \to B}(\omega) = \frac{I_{B,\mathbf{n}}(\omega)}{I_{A,\mathbf{n}}(\omega) + I_{B,\mathbf{n}}(\omega)} \tag{8}$$

where $I_{A/B,\mathbf{n}}(\omega)$ is expressed by:

$$I_{A,\mathbf{n}}(\omega) = \frac{1}{2} \cdot \frac{\Omega}{(2\pi)^3} \sum_{states\ s\ in\ BZ_A} \delta(\omega_s - \omega) |\mathbf{v}_s \cdot \mathbf{n}| \tag{9}$$

Where $\mathbf{n}$ is a unit vector perpendicular to the interface. The summation is performed over all the phonon states *s* in the materials *A* in its entire first Brillouin zone $BZ_A$.

Hence, when a phonon collides with the interface, one of the reflection (phonon remaining in the same material) or transmission (phonon transferred to the other side of the interface) process is selected randomly according to the transmission probability given by Eq.(9). Then, the wave vector and the mode of the final phonon are randomly selected (independently of the incident state) among the isoenergy states *j* in the final material with a non-uniform probability $P_j$ to satisfy the Lambert's cosine law (cf. Eq. 7) i.e.: $P_j(\mathbf{q}_{j\prime}, \theta_r) \propto |\mathbf{v}_{j\prime}(\mathbf{q}_{j\prime})| \cos \theta_r$.

### II.5. Investigated structures

In this work, homogeneous and heterogeneous nanostructures of length *L* along the X axis were investigated. In all structures, the thermal flux flows along the X axis. The cells located at the two extrema have their external face placed in contact with a hot thermostat of temperature $T_H$ and a cold thermostat of temperature $T_C$, respectively.

All nanostructures have square cross-sections. Their width and height are equal to 100 nm. A cubic mesh is employed. First, all structures are uniformly meshed along the X-axis into 20 equally sized cells. Then, the meshing of the heterojunctions is refined in the vicinity of the interfaces and the thermostats. The length of these refined cells located 5 nm around the interfaces are less or equal to 1 nm.

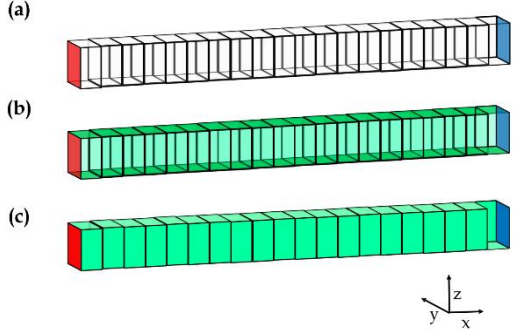 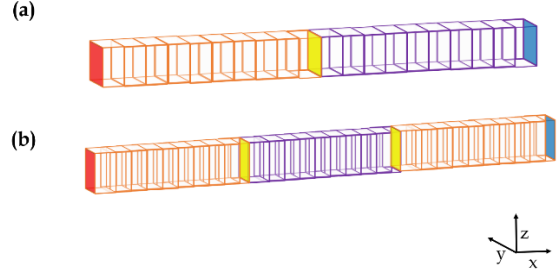

*Fig. 3. Homogeneous nanostrucrures: (a) Cross-plane nanofilm (CPNF). (b) In-plane nanofilm (IPNF). (c) Nanowires (NW). Red/blue faces are hot/cold thermostats, respectively. Transparent external faces: periodic boundaries. Green faces: rough boundaries.*

*Fig. 4. Heterogeneous nanostructures: (a) Simple heterojunction. (b) Double heterojunction. Red/blue faces are hot/cold thermostats. Transparent external faces: periodic boundaries. Yellow faces: DMM interfaces.*

Three configurations of homogeneous nanostructures illustrated in Fig. 3 are studied:

- Cross-plane nanofilms (CPNF) in which periodic boundary conditions are applied along both Y and Z directions to describe semi-infinite boundaries.
- In-plane nanofilms (IPNF) in which periodic boundary conditions are applied for XZ planes to describe semi-infinite boundaries and diffusive boundary conditions are applied to XY planes to represent external faces (rough boundaries).
- Nanowires (NW) in which the diffusive boundary conditions are applied to all XY and XZ planes.

In addition, two different kinds of heterogeneous nanostructures in the cross-plane configuration schematized in Fig. 4 are studied:

- Simple heterojunctions in which a DMM heterojunction (in yellow) is located in the middle of the structure ($x = L/2$) joining two homogeneous bars.
- Double heterojunctions in which three equally long homogeneous bars are joined to form two DMM interfaces at positions $x = L/3$ and $x = 2L/3$.

## III  Results

### III.1. Ge homogeneous nanostructures

As shown previously for Si in Ref [34], Fig. 5 shows the variation of the thermal conductivity of Ge in CPNF, IPNF and NW as a function of the nanostructure length. When the nanostructure length $L$ becomes smaller than a few tens of nanometers which is the order of magnitude of the effective phonon mean free path in these structures, the phonon transport regime tends to be more and more ballistic. For the three nanostructures, the thermal conductivity follows the same behavior, i.e. it tends asymptotically to the ballistic thermal conductivity of Ge (dashed line) that is a linear function of *L*. In long nanostructures, i.e. much longer than the phonon mean free path, the transport regime becomes diffusive, and the thermal conductivity tends to be *L* independent in all structures. In CPNF, the thermal conductivity moves towards the experimental conductivity in bulk Ge of 60 Wm$^{-1}$K$^{-1}$ at 300 K [46] and its evolution is consistent with *ab-initio* results from [47] and agrees reasonably with the experimental evolution presented in Ref [48]. In nanostructures IPNFs and NWs, the rough interfaces contribute to reduce the thermal conductivity at large *L*. Consistently, for a given system length, NWs exhibit a lower conductivity than IPNFs that have a lower conductivity than CPNFs.

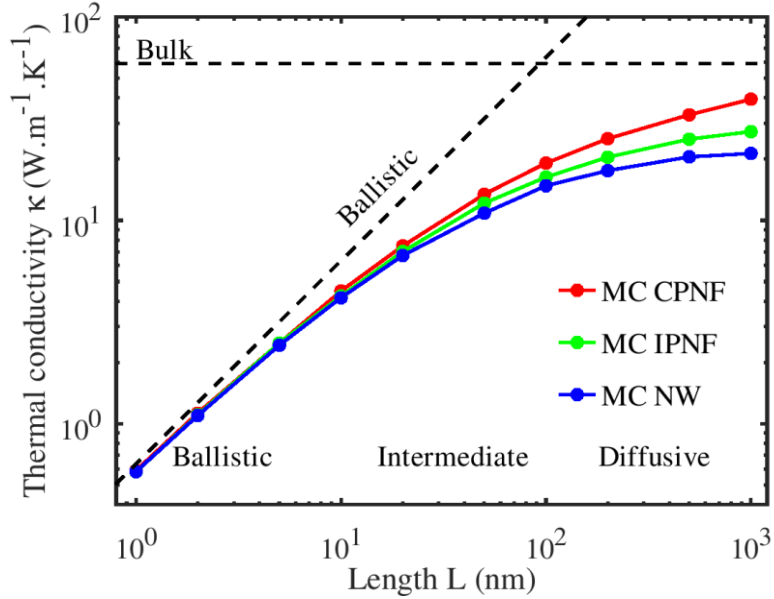

*Fig. 5. Thermal conductivity of Ge homogeneous nanostructrues of the three homogeneous nanostructures as a function of the nanostructure length L.*

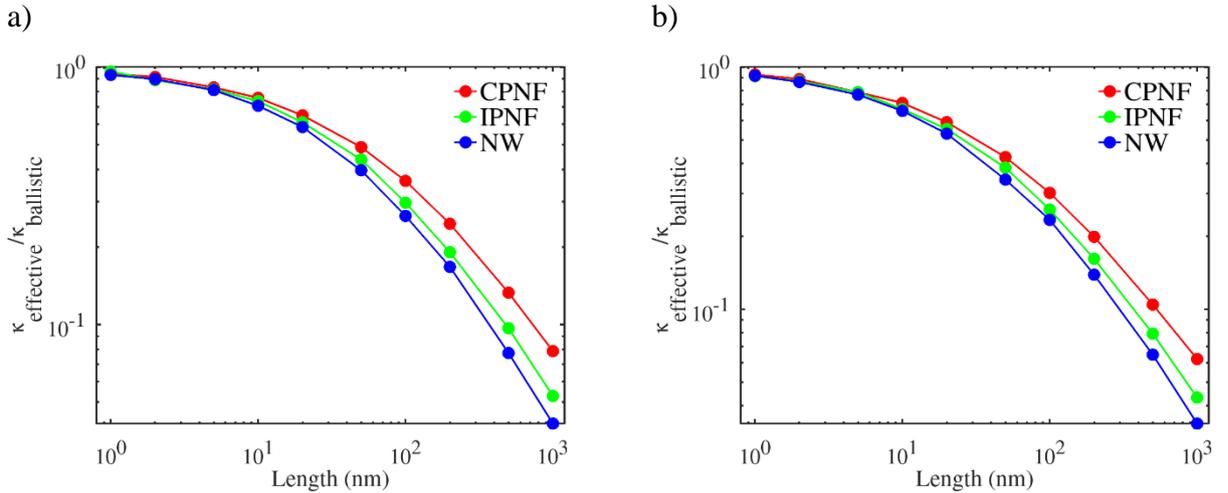

*Fig. 6. The ratio $\kappa_{effective}/\kappa_{ballistic}$ as a function of the length L. (a) Si nanostructures (b) Ge nanostructures.*

To quantify the transition between the different phonon transport regimes, the evolution of the Knudsen number $K_D$ is widely used. This number, characterizing the degree of ballisticity of transport, is defined as the ratio between the effective conductivity $\kappa_{effective}$ (see Eq. 3) and the ballistic one $\kappa_{ballistic}$ (see Eq. 4). It depends on the nanostructures as follows:

$$K_D = \kappa_{effective}/\kappa_{ballistic} = \begin{cases} \kappa_{CPNF}/\kappa_{ballistic} & for\ CPNF \\ \kappa_{IPNF}/\kappa_{ballistic} & for\ IPNF \\ \kappa_{NW}/\kappa_{ballistic} & for\ NW \end{cases} \quad (10)$$

The values of $K_D$ obtained from our MC code are plotted in Fig. 6 for different nanostructures of length $L$ varying from 1 nm to 1 µm made of Si (a) and Ge (b).

As expected, the value of $K_D$ decreases monotonically from 1 in ultra-short systems where a fully ballistic transport regime occurs down to 0 (i.e. smaller than 0.1) in long systems in which the transport is fully diffusive. The intermediate transport regime is defined here as the regime where $K_D$ values are comprised between 0.8 ($L > 10$ nm) and 0.1 ($L < 1$µm).

Beyond the Knudsen number, Monte Carlo simulations can provide a deep insight into the phonon transport as they give access to the spectral phonon distributions in nanostructures. In Fig. 7. Variation of modal contribution of modes to the total heat flux as a function of the CPNF length: (a) in Ge; (b) in Sithe contribution of each phonon mode to the total heat flux is plotted in percentage for different CPNFs of different lengths made of Si (a) and Ge (b). Interestingly, these spectral contributions are different when the device length differs, i.e. when the transport regime changes.

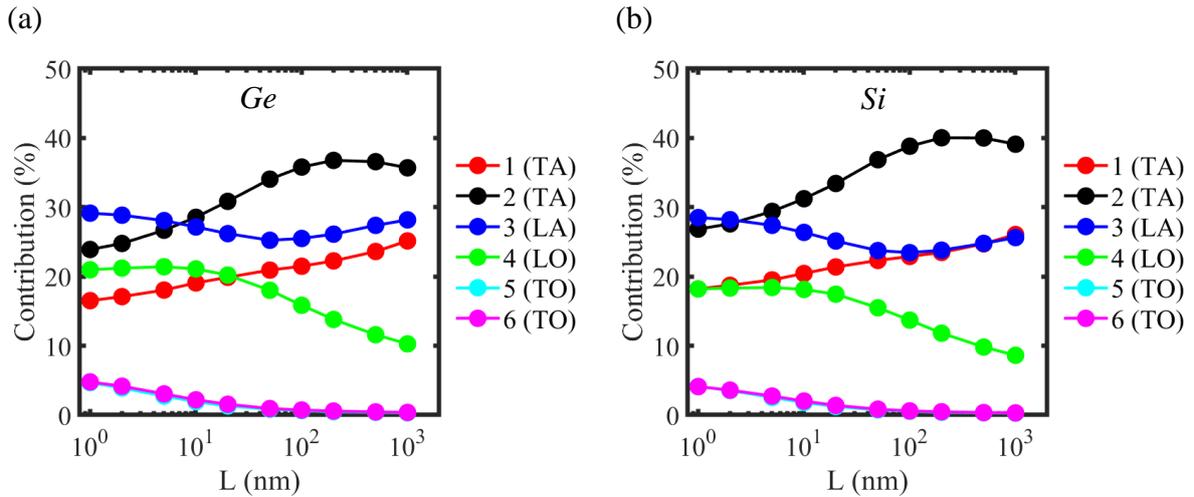

Fig. 7. Variation of modal contribution of modes to the total heat flux as a function of the CPNF length: (a) in Ge; (b) in Si

Therefore, in the following we will investigate how these modal contributions can be used to characterize the transport regime in nanostructures. For this purpose, the monotonous evolutions of the optical mode contributions in both Ge and Si is interesting. The main advantage of this (indirect) indicator is that it is locally defined while $K_D$ is a "macroscopic" indicator.

### III.2. Simple Si/Ge heterostructure

Fig. 8 shows the profiles of the local pseudo-temperature $T$ and the two hemispherical temperatures $T^+$ and $T^-$ in a 100 nm long Ge/Si heterojunction. The Ge bar is in contact with a hot thermostat at temperature $T_H$ = 302 K and the Si bar is in contact with a cold thermostat at temperature $T_C$ = 298 K. Along the two homogenous bars, the different temperatures vary continuously whereas they exhibit an abrupt drop at the interface. More surprisingly, all the considered temperatures do not evolve strictly linearly along the bars but have a significant curvature near the interface, particularly in the Ge side that has a lower thermal conductivity than Si. Thus, a coarse mesh around the interface could lead to a significant overestimation of the temperature drop at the interface and an underestimation of the related interface thermal conductance. For this reason, the refinement of the mesh in the vicinity of the interfaces is mandatory.

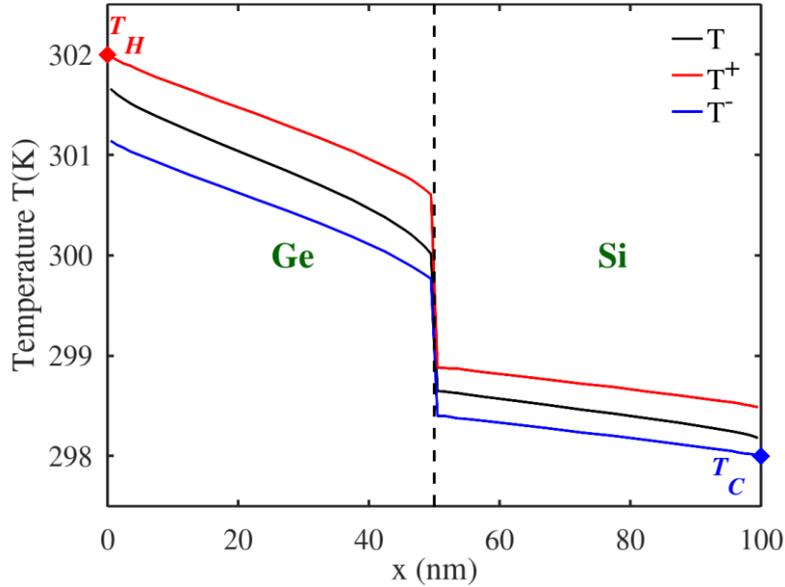

*Fig. 8. Temperature profiles in a 100 nm long Ge/Si heterojunction*

For a 20 nm long Si/Ge heterojunction, the heat flux resulting from Monte Carlo simulation is plotted in Fig. 9.*a* for different temperature differences *($T_H$-$T_C$)* between thermostats while keeping the average temperature at 300 K. The Si/Ge heterojunction was considered for both configurations: (i) $T_H$:Ge/$T_C$:Si in which the Ge bar is in contact with the hot thermostat while the Si bar is in contact with the cold thermostat; and (ii) $T_H$:Si/$T_C$:Ge in which the Si bar is in contact with the hot thermostat while the Ge bar is in contact with the cold thermostat.

Fig. 9.*a* clearly shows a linear relationship between the heat flux density and the temperature difference *($T_H$-$T_C$)* for both thermal biases. By using linear regressions, the following relationships are obtained:

i. $T_H$:Ge/$T_C$:Si : $Q = (191.30 \text{ MW.m}^{-2}.\text{K}^{-1})(T_H - T_C) + 14.40 \text{ MW.m}^{-2}$
ii. $T_H$:Si/$T_C$:Ge : $Q = (191.35 \text{ MW.m}^{-2}.\text{K}^{-1})(T_H - T_C) - 16.24 \text{ MW.m}^{-2}$

A total thermal conductance of 191 MW.m$^{-2}$.K$^{-1}$, independent of the temperature difference configuration, is thus obtained in these 20 nm–long heterojunction. Due to the numerical nature of the Monte Carlo method, the heat flux density at zero temperature bias does not vanish, but it takes a finite value around -/+16 MW.m$^{-2}$ in this device. We thus assume that this value corresponds to the maximum numerical resolution $\delta Q$ of our simulation. Besides, in a MC algorithm the interval of 95% confidence of the flux called $IC_{95}Q$ can be reduced as small as expected if the number of samples is large enough. Finally, the error bars for heat flux density are defined here as $Errorbar_Q = min\{\delta Q, IC_{95}Q\}$. Similarly, the error bars for the interface thermal conductance are defined by summing:

$$\frac{\delta G_{int}}{G_{int}} = \frac{\delta Q}{Q} + \frac{\delta(\Delta T^I)}{\Delta T^I} \qquad (11)$$

where our estimation of $\delta(\Delta T^I)$ is $\delta T \approx \delta Q/G_{tot}$.

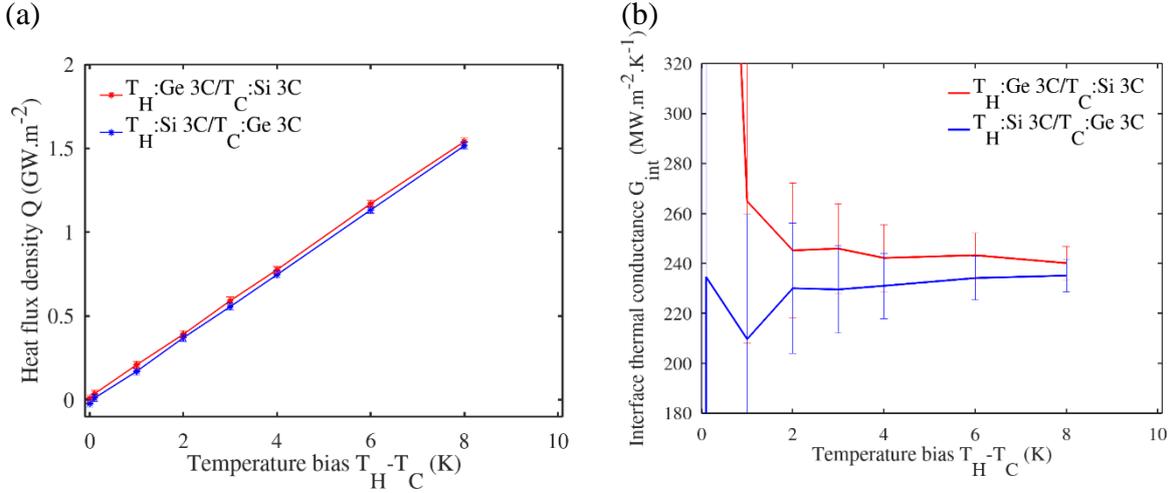

*Fig. 9. (a) Heat flux density and (b) interface thermal conductance in a 20 nm – long Si/Ge heterojunction for different temperature differences ($T_H$-$T_C$). N.B.: ($T_H$+$T_C$)/2=300 K*

Fig. 9.*b* presents the interface thermal conductance as a function of the temperature difference ($T_H$-$T_C$). For a temperature difference greater than or equal to 2K, the interface thermal conductance converges to a value of about 240 MW.m$^{-2}$.K$^{-1}$ (or a resistance of 4.2×10$^9$ K.m/W) that is consistent with the experimental value of Refs. [27], [45], [49]. Although the error bars are large at low temperature bias, due to the numerical fluctuations inherent in the Monte Carlo method, the value of $G_{int}$ is still within the error bars, indicating that $G_{int}$ is independent on the temperature difference ($T_H - T_C$). In Ref. [50] a rectification of 5% was estimated by using MD at Si/Ge interfaces. This effect could be due to (i) the asymmetric phonon distributions on each side of the interface, phenomenon captured by our approach, and/or (ii) the inelastic scattering at the interfaces that is not included here. As there is an overlap of the error bar extension in the two thermal configurations, it is difficult to conclude here. Nevertheless, a small thermal rectification of less than 5% is compatible with the results observed in Fig. 10.b.

A set of Si/Ge heterojunctions of length *L* varying from 1 nm to 1 µm with thermostat temperatures of $T_H$ = 302 K and $T_C$ = 298 K have been also considered. The heat flux density of the heterojunctions as a function of the heterostructure length *L* is plotted Fig. 10.*a*. In ultra-short heterostructures (*L*<10 nm), the heat flux density *Q* approaches the value of 1 GWm$^{-2}$ in ballistic regime. In long heterostructures (*L*>200 nm) in which the heat transport is diffusive, the heat flux density tends to be proportional to $1/L$ consistently with the classical Fourier's law. The related thermal conductances $G_{int}$ are plotted in Fig. 10.b. As the values of $G_{int}$ remains within the error bars, the interface thermal conductance takes average values close to 240 MWm$^{-2}$K$^{-1}$ for both temperature difference configurations and for all lengths $L$. Once again, noted that the width of the error bars is larger for long heterostructures (in which $\Delta T^I$ of Eq. 11 are low) than for short ones.

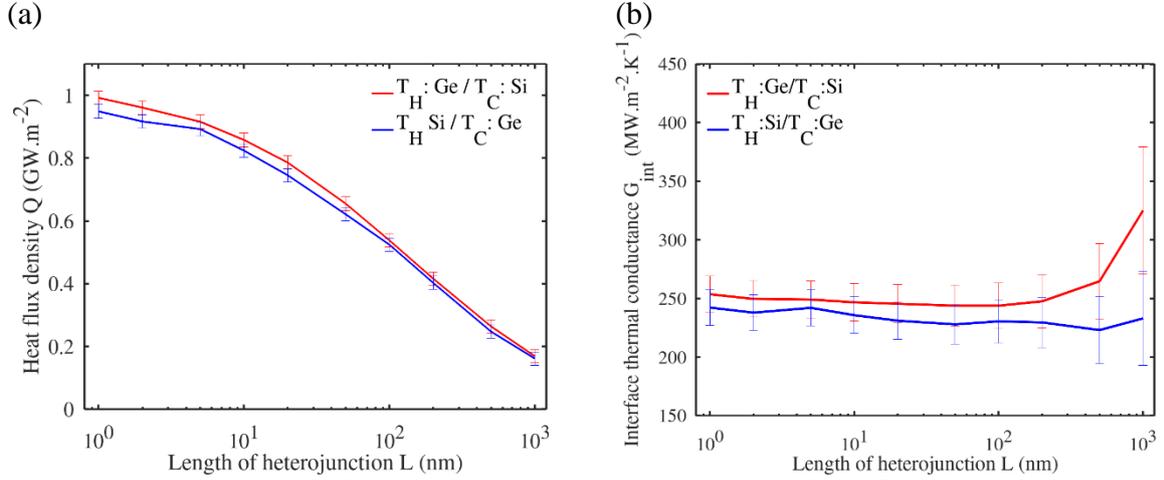

*Fig. 10. (a) Heat flux density and (b) interface thermal conductance of Si/Ge heterojunction as a function of the heterostructure length L ($T_H$=302 K, $T_C$=298 K).*

Another interesting thermal parameter of an interface is its Kapitza length $l$ defined as $l=\kappa_{diffusive}/G_{int}$, where $\kappa_{diffusive}$ is the thermal conductivity. It corresponds to the length of a homogeneous bar made of material exhibiting the same thermal conductance as one interface. Here, the Kapitza lengths are 540 nm and 250 nm in Si and Ge, respectively. In long device, longer than the Kapitza length of the interface, the presence of the interfaces becomes negligible and the heat flux, presented in Fig. 10.a, tends to evolve as $1/L$. In short devices, it tends to be size independent. Indeed, the interface conductance limits the flux and not the ballistic one as in homogenous devices.

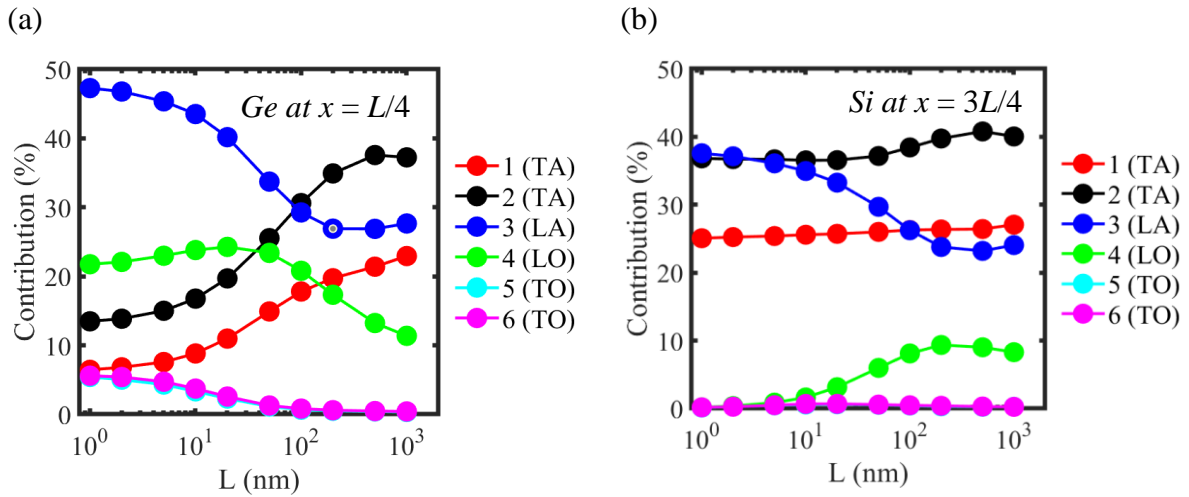

*Fig. 11. Variation of modal contribution of modes in $T_H$:Ge/$T_C$:Si heterojunctions as a function of the heterostructure length: (a) in the Ge side at x=L/4; (b) in the Si side at x=3.L/4.*

To investigate spectrally the phonon transport in heterostructures, the contribution to the total heat flux density of all the phonon modes in the middle of each side of the heterostructure, i.e. at *x=L*/4 in Ge (a) and at *x=3.L*/4 in Si (b), is shown in Fig. 11. Variation of modal contribution of modes in $T_H$:Ge/$T_C$:Si heterojunctions as a function of the heterostructure length: (a) in the Ge side at x=L/4; (b) in the Si side at x=3.L/4. as a function of the nanostructure length *L*.

Interestingly, the evolutions of the modal contributions in the Ge and Si parts of the heterostructure (Fig. 11. Variation of modal contribution of modes in $T_H$:Ge/$T_C$:Si heterojunctions as a function of the heterostructure length: (a) in the Ge side at x=L/4; (b) in the Si side at x=3.L/4..a) are in overall accordance with the results obtained in homogenous CPNF previously shown in Fig. 7. Variation of modal contribution of modes to the total heat flux as a function of the CPNF length: (a) in Ge; (b) in Si Indeed, the evolution of this mode in Si CPNF and that in Si/Ge heterostructure have inverse evolution as their contribution is reduced when the length is reduced. This is due to the fact that the optical modes cannot contribute to the heat flux in Silicon near the Si/Ge interface due to phonon mode mismatch between Si and Ge. This mismatch does not appear at the interface Si/contact present in CPNF as this interface is assumed to be perfectly transparent..

Thus, the modal contributions are fixed when the transport is near equilibrium or diffusive. However, since different out of equilibrium regimes can occur according to the type of boundary conditions, different modal contributions can be observed in such regimes.

Finally, the two-transverse optical (TO) mode in Ge remains of special interest in all cases as their contribution to the total heat flux decreases monotonically when increasing the length *L* of the nanostructure and a contribution above 1% is a signature of a strongly out of equilibrium transport regime. For silicon we have to compare the modal contributions close and far from the interfaces.

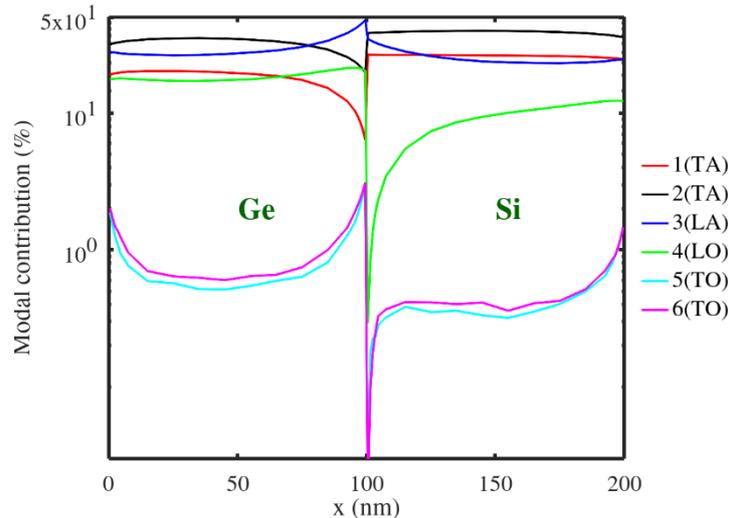

*Fig. 12. Evolution of modal contribution to total heat flux density in 200 nm – long simple heterojunction $T_H$: Ge/$T_C$: Si.*

To illustrate the statements, the spatial evolution of modal contributions to the total heat flux along the X axis is plotted in Fig. 12. Evolution of modal contribution to total heat flux density in 200 nm – long simple heterojunction $T_H$: Ge/$T_C$: Si. As mentioned above, in the middle of the Ge and Si bar all modal contributions are distributed according to Fig. 11. Variation of modal contribution of modes in $T_H$:Ge/$T_C$:Si heterojunctions as a function of the heterostructure length: (a) in the Ge side at x=L/4; (b) in the Si side at x=3.L/4.. For instance, the TO contributions are smaller than 1% which correspond to same percentage as that observed in long device with L > 200 nm associated with a $K_D$ < 0.1, i.e. in the case of a diffusive transport regime.

Near the interfaces, i.e. near the Si/Ge interface and the contact interfaces, modal contributions are significantly different from the equilibrium distribution indicating the occurrence of strongly out of equilibrium transport regimes. In Ge, the TO contribution is higher than 2% as in short bar where the transport is quasi ballistic. Differently, in the Si part, we observe two different evolutions of the TO contribution: near the Si/Ge interface where the TO contribution is reduced and near the contact where the TO contribution is enhanced. Near an interface the modal contributions are directly related to the type of this interface: ideal emitter as at the contact or diffusive as at the Si/Ge interface.

### III.3. Double Si/Ge heterostructures

To study more complex systems, several heterostructures with two Si/Ge interfaces, schematized in Fig. 4.b, have been investigated. The temperature profiles in a 300 nm long Ge-Si-Ge double heterojunction are plotted in Fig. 13.a. The temperature profiles in both Ge parts are similar as well as the temperature drops at the two interfaces. As previously observed in simple heterostructures, the temperature gradients are smaller in the more conductive Si part and significant curvatures of the temperature profiles are present in the vicinity of the interface, in particular on the Ge side. In addition, the *T* gradient is always slightly higher than the *T*⁺ and *T*⁻ ones. The spatial evolution of modal contributions to heat flux along the X axis is plotted in Fig. 13.b. In a double heterostructure, these evolutions are similar to those occurring in a simple Si/Ge heterostructure. The modal contributions are perturbed by the interface and differ from that occurring in the central region where a distribution close to equilibrium is recovered. The phonon transport regime near the interfaces appears to be significantly out of equilibrium.

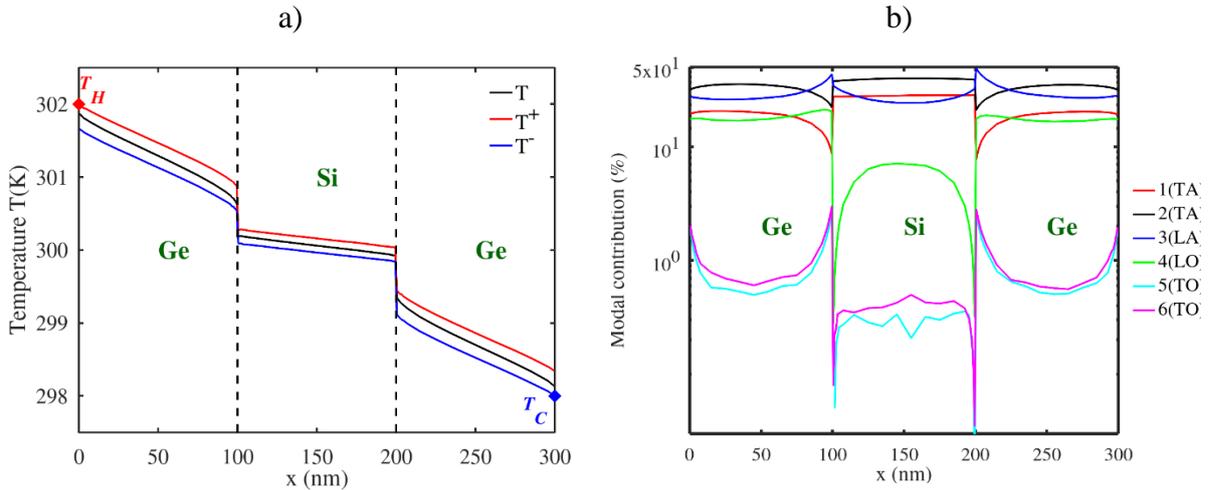

*Fig. 13. Profiles of a) temperatures and b) modal contributions to total heat flux density in a 300 nm - long double heterojunction Ge/Si/Ge*

According to our Monte Carlo results, the total thermal flux density flowing through the double heterojunction is 240±21 MWm⁻². The calculated values for the two interface thermal conductances are $G_1 = 271\pm32$ MWm$^{-2}$K$^{-1}$ (at $x = 100$ nm) and $G_2 = 235\pm28$ MWm$^{-2}$K$^{-1}$ (at $x = 200$ nm). With respect to the value extracted in simple heterostructures (of 240 MWm$^{-2}$K$^{-1}$), the values obtained for double heterostructures remain within the error bars of the method.

In addition, the effective thermal conductivities of the Ge and Si bars are $17.3\pm1.7$ Wm$^{-1}$K$^{-1}$, $46.9\pm5.9$ Wm$^{-1}$K$^{-1}$ and $17.3\pm1.7$ Wm$^{-1}$K$^{-1}$, for the left Ge bar ($0 \leq x \leq 100$ nm), the middle Si bar ($100$ nm $\leq x \leq 200$ nm), and the right Ge bar ($200$ nm $\leq x \leq 300$ nm), respectively. These results match the values obtained in 100 nm long CPNF and shown in Ref [34] and Fig. 5, respectively. They illustrate the relevance of our simulation methodology even in the presence of several interfaces.

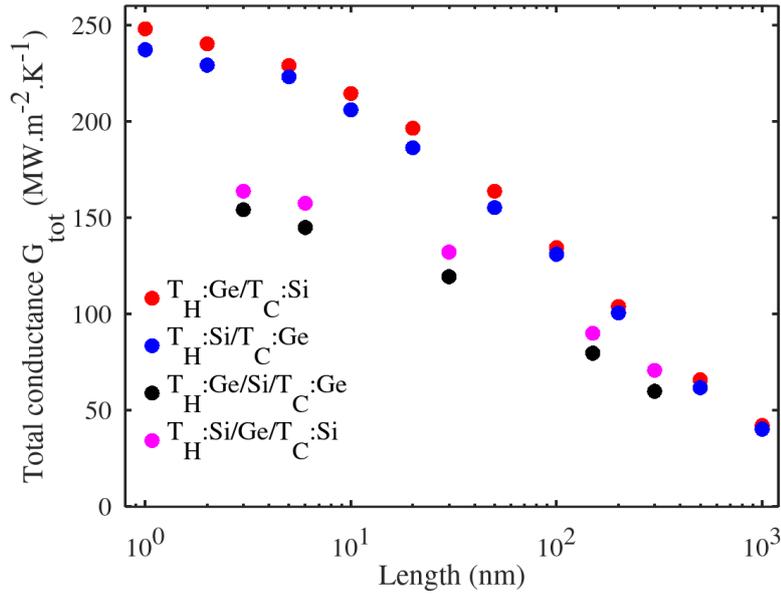

Fig. 14. Total conductance of simple and double Si/Ge heterojunctions cf. Fig 4 of various lengths L at T=300 K.

It is important to note that this consistency of MC results in terms of interface conductance and thermal conductivity has been obtained thanks to the refinement of the space meshing near the interface associated with the use of the hemispherical temperatures. Otherwise, the computed interface thermal conductance would appear artificially size dependent (not shown).

Finally, in Fig. 14, the total conductance of the simple and double heterostructures computed by using our MC simulations, is plotted as a function of the device length $L$. In ultra-small devices working in ballistic transport regime, the total conductance in heterostructures tends to become size independent and governed by the Si/Ge interface conductance. Indeed, in the single ultra-short heterostructures, we find a thermal conductance value of about 240 MW/m$^2$/K which corresponds to that of a Si/Ge interface. It is interesting to note that in the double ultrashort heterostructures, the value of the total conductance is higher than those of the single heterostructure divided by two, which would be the result if we had just two interface conductances in series. This indicates that when the interfaces are too close together, they are less effective at stopping heat transport, because the phonon transport is far from equilibrium between the two interfaces.

In long device, much longer than the Kapitza length of the interface, the presence of the interfaces becomes negligible. The main advantage of the presented MC approach lies in its ability to describe the heat transfer in the intermediate regime even in complex geometries such as double heterostructures of few tens of nanometers length.

# IV    Conclusion

A Full Band Monte Carlo simulator able to consider phonon transmission across interfaces disposed perpendicularly to the heat flux has been presented. In the present work, the phonon transmissions were computed by using a Full band version of the DMM formalism.

Simulations have been used to study single and double heterostructures made of Si/Ge interfaces. A clear methodology has been developed to extract relevant results from the numerical output data. By considering the hemispherical temperatures, i.e. by distinguishing the phonons by the direction of their displacement and by carefully refining the mesh around the interfaces to properly capture the curvature of the temperature profiles, we have calculated accurate values of thermal interface conductance that are not size dependent or heterostructure dependent, as expected.

As allowed by the Monte Carlo algorithm, the spectral contribution to the thermal flux of the different phonon modes has been highlighted and correlated with the phonon transport regime. In particular, the out-of-equilibrium transport regime occurring in the vicinity of interfaces has been clearly demonstrated and illustrated.

This original Monte Carlo simulator providing a deep insight of the phonon transport can be used to study complex geometries with many interfaces of different types: rough/specular, reflecting, or semi-transparent interfaces. In addition to the spectral analysis of phonon transport [38], it opens the way to the investigation of the transient aspects of heat conduction in semiconductor devices.

## Acknowledgements

This work was supported by a public grant overseen by the French National Research Agency (ANR) as part of the "Investissements d'Avenir" program (Labex NanoSaclay, reference: ANR-10-LABX-0035).

## Data availability

The data that support the findings of this study are available from the corresponding author upon reasonable request.